\documentclass[12pt]{article}
\begin{document}
\begin{center}
\Huge{Mass Shifts through Re-scattering}\\
\vspace {.2in}
\Large{Ron  S. Longacre}\\
\large{\it Brookhaven National Laboratory, Upton, NY 11973, USA\footnote{
This research was supported by the U.S. Department of Energy under Contract No. DE-AC02-98CH10886}}\\
\vspace{.5in}
\end{center}
\begin{abstract}
In this note we present a model that can produce a mass shift
in a resonance due to interference between a scattering amplitude
and that amplitude having rescattering through the resonance.
\end{abstract}
\section{Starting point}
For the first part of this story we will define what a scattering
cross section is. We will only consider in this work elastic
scattering of pions. Two pions can scatter at a certain energy
which we will call $M_{\pi \pi}$. The differential cross section
$\sigma$ at a given $M_{\pi \pi}$ is
\begin{equation}
\frac{d\sigma}{d\phi d\theta} = \frac{1}{K^2} \left| \sum_{\ell}
(2\ell + 1) T_{\ell} P_{\ell} (cos(\theta)) \right| ^2
\end{equation}
where $\phi$ and $\theta$ are the azimuthal and scattering angles,
respectively. $T_{\ell}$ is a complex scattering amplitude and
$\ell$ is the angular momentum. $P_{\ell}$ is the Legendre
polynomial, which is a function of $cos(\theta)$. $K$ is the flux
factor equal to the pion momentum in the center of mass. For this
note we only consider $\ell =$ 0 and $\ell =$ 1. The $T_0$ and
$T_1$ elastic scattering amplitudes are complex amplitudes
described by one real number which is in units of angles. The form
of the amplitude is
\begin{equation}
T_{\ell} = e^{i \delta_{\ell}} sin(\delta_{\ell})
\end{equation}
We note that $\delta$ depends on the value of $\ell$ and $M_{\pi
\pi}$. We will use the $\ell =$ 0 and $\ell =$ 1 $\delta$'s given
in \cite{1}.
\section{Defining A}
Let us consider two pions scattering in the final state of the
heavy ion collision. The scattering will be either $\ell =$ 0 or
$\ell =$ 1 partial waves. The $M_{\pi \pi}$ of the scattering
di-pion system will depend on the probability of the phase space
of the overlapping pions. The pions emerge from a close encounter
in a defined quantum state with a random phase. We will call this
amplitude $A$ and note that the absolute value squared of the
amplitude is proportional to the phase space overlap. The emerging
pions can re-interact or re-scatter through the quantum state
of the pions, which is a partial wave or a phase shift. We have
amplitude $A$ plus $A$ times the re-scattering of pions through
the phase shift consistent quantum state of $A$. The correct
unitary way to describe this process is given by \cite{2} equation
(4.5)
\begin{equation}
T = \frac{V_1 U_1}{D_1} + \frac{\left( V_2 + \frac{D_{12}
V_1}{D_1} \right)\left( U_2 + \frac{D_{12} U_1}{D_1}\right)}{D_2 -
\frac{D_{12}^2}{D_1}}
\end{equation}
In the above equation we have two terms, 1 and 2. The first term 
denoted by 1 is the $\pi \pi$ scattering through p-wave which will 
become the amplitude $A$ mentioned above, where $V$ is the incoming 
and $U$ is the outgoing $\pi \pi$ system. The second term denoted 
by 2 is the direct production of the $\rho$ meson with $V$ being 
the production, the propagation being $D$ and the decay being $U$. 
We see that there are terms $D_{12}$ which involves a loop of pions 
between scattering pions and the formation of a $\rho$ by the pions. 
\section{Final equation}
The complete derivation is in the appendix. From the appendix we 
get two terms, one being the direct production of the $\rho$ or 
$\pi \pi$ p-wave phase shift and the second being the $\rho$ from 
re-scattering. The final equation number 21 has two important factors,
one is two-body phase space and the other is a coefficient $\alpha$.
This coefficient is related to the real part of the $\pi \pi$ rescattering
loop and is given by equation 14. When the pions rescatter or interact at
a close distance or a point $\alpha$ has its maximum value of one. While
if the pions rescatter or interact at a distance determined by the diffractive
limit the value of $\alpha$ is zero.

In the equation 21 $|T|^2$ is the cross section for $\rho$
production, where $D$ is the direct production amplitude and $A$
is the amplitude introduced above for the re-scattering pions into
the $\rho$ meson. $\delta$ is the $\pi \pi$ phase shift \cite{1}.
$q$ is the $\pi \pi$ center of mass. At a given $p_T$ and $y$ bin,
$D$ will be a constant as a function of $M_{\pi \pi}$. The $\alpha$
factor is one for rescattering coming from a point source, but
goes to zero when rescattering is diffractive. The dependence of 
$A$ is calculated by the phase space overlap of di-pions added 
as four vectors and corrected for proper time, with the sum 
having the correct $p_T$ and $y$ for a given $M_{\pi \pi}$.

Finally we most use the correct two body phase space. For a two body
system of pions, phase space goes to an constant as $M_{\pi \pi}$ goes
to infinity. Let us choose this constant to be unity. Phase space which
is denoted by PS is equal to
\begin{equation}
PS = \frac{2qB_{\ell}(q/q_0)}{M_{\pi \pi}}
\end{equation}
where $B_{\ell}$ is a Blatt-Weisskopf-barrier factor\cite{3} for $\ell$ 
angular momentum quantum number. The $q_0$ is the momentum related to the 
range of interaction of the $\pi \pi$ scattering. 1 fm is the usual 
interaction distance which implies that $q_0$ is 200 MeV/c. For the $\rho$ 
meson $\ell$=1 and $B_1$ = $\frac{(q/q_0)^2}{(1+(q/q_0)^2)}$.  

The phase space factor PS as a function of $q$ near the $\pi \pi$ threshold 
is given by $q^{2\ell + 1}$. Thus in the appendix we use $q^{2\ell + 1}$ for
the factor PS except for equation 21 which is the final equation. 

\appendix
\section{Appendix}

Starting with equation (4.5) from \cite{2}
\begin{equation}
T = \frac{V_1 U'_1}{D_1} + \frac{\left( V_2 + \frac{D_{12}
V_1}{D_1} \right)\left( U'_2 + \frac{D_{12} U'_1}{D_1}\right)}{D_2
- \frac{D_{12}^2}{D_1}}
\end{equation}
In order to have the correct threshold kinematics, we define
\begin{equation}
U'_1 = U_1 \sqrt{q^{2\ell + 1}}
\end{equation}
\begin{equation}
U'_2 = U_2 \sqrt{q^{2\ell + 1}}
\end{equation}
where $q$ is the $\pi \pi$ center of mass momentum and $\ell$ is
the value of the angular momentum. The amplitude $A$ of the text
is given by
\begin{equation}
\frac{V_1 U_1}{D_1} = A
\end{equation}
Thus we have
\begin{equation}
\frac{V_1 U'_1}{D_1} = A \sqrt{q^{2\ell + 1}}
\end{equation}
The phase shift for the $\ell^{th}$ partial wave will be given by
$\delta_\ell$, where
\begin{equation}
\frac{U'_2 U'_2}{D_2} = e^{i \delta_\ell} sin(\delta_\ell)
\end{equation}
The above equality is true if the $D_1$ mode plays no role in the
$\pi \pi$ scattering in the $\ell^{th}$ partial wave. But in the
initial state there is a large production of $D_1$. The $U'$s are
the basic coupling of the $D'$s to the $\pi \pi$ system. In order
to decouple $D_1$ from the $\pi \pi$ system $U_1$ must go to zero.
We can maintain a finite production of $D_1$ if we define
\begin{equation}
V_1 = \frac{1}{U_1}
\end{equation}
Thus the first term in the equation becomes
\begin{equation}
\frac{V_1 U'_1}{D_1} = \frac{\frac{1}{U_1} U_1 \sqrt{q^{2\ell +
1}}}{D_1} = \frac{\sqrt{q^{2\ell + 1}}}{D_1} = A \sqrt{q^{2\ell +
1}}
\end{equation}
The form of $D_{12}$ is given by
\begin{equation}
D_{12} = \alpha U_1 U_2 + i q^{2\ell + 1} U_1 U_2
\end{equation}
$D_{12}$ is the real and imaginary part of the two pion loop from
state 1 to state 2. The $U'$s are the $\pi \pi$ couplings and the
imaginary part goes to zero at the $\pi \pi$ threshold. The $\alpha$
factor is one for rescattering coming from a point source, but
goes to zero when rescattering is diffractive. A simple form for $\alpha$
is given by 
\begin{equation}
\alpha = (1.0 - \frac{r^2}{r_0^2}) 
\end{equation}
where $r$ is the radius of rescattering in fermis and $r_0$ is 1.0
fermi or the limiting range of the strong interaction.
The second term of the first equation is
\begin{equation}
\frac{\left( V_2 + \frac{D_{12} V_1}{D_1}\right) \left( U'_2 +
\frac{D_{12} U'_1}{D_1}\right)}{D_2 - \frac{D_{12}^2}{D_1}}
\end{equation}
Rewriting
\begin{equation}
\frac{\left( V_2 + \frac{\alpha U_1 U_2 V_1}{D_1} + i q^{2\ell + 1}
\frac{U_1 U_2 V_1}{D_1}\right) \left( U'_2 + \frac{D_{12}
U'_1}{D_1} \right)}{D_2 - \frac{D_{12}^2}{D_1}}
\end{equation}
Let us make substitutions
\begin{equation}
V_1 = \frac{1}{U_1}, U_2 = \frac{U'_2}{\sqrt{q^{2\ell + 1}}} ,
\frac{1}{D_1} = A, D_{12} = 0
\end{equation}
The second term becomes
\begin{equation}
\frac{\left( V_2 + \frac{A \alpha U'_2}{\sqrt{q^{2\ell + 1}}} + i
\sqrt{q^{2\ell + 1}} A U'_2 \right) U'_2}{D_2}
\end{equation}
The first term is
\begin{equation}
\frac{V_1 U'_1}{D} = A \sqrt{q^{2\ell + 1}}
\end{equation}
Adding the first and the second terms and substituting the phase
shift,
\begin{equation}
T = \frac{V_2}{U_2} \frac{e^{i \delta_\ell}
sin(\delta_\ell)}{\sqrt{q^{2\ell + 1}}} + A \left( \frac{e^{i
\delta_\ell} \alpha sin(\delta_\ell)}{\sqrt{q^{2\ell + 1}}} +
\sqrt{q^{2\ell + 1}} e^{i \delta_\ell} cos(\delta_\ell) \right)
\end{equation}
The term with the factor $\frac{V_2}{U_2}$ is the direct
production of the di-pion system. We shall call this amplitude
$D$. The re-scattered amplitude is $A$ and is modified by the
di-pion phase shift. These two amplitudes have some random phase
and are not coherent. Thus the cross section is
\begin{equation}
|T|^2  = |D|^2 \frac{sin^2(\delta_{\ell})}{PS} +
\frac{|A|^2}{PS} \left| \alpha sin(\delta_\ell) + PS cos(\delta_\ell) \right|^2
\end{equation}


\begin{thebibliography}{9}
\bibitem{1} G. Grayner {\it et. al.}, Nucl. Phys. B 75 (1974) 189.
\bibitem{2} R. Aaron and R. S. Longacre, Phys. Rev. D 24 (1981)
1207.
\bibitem{3} F. von Hippel and C. Quigg, Phys. Rev. 5 (1972) 624.
\end{thebibliography}
\end{document}